\documentclass[superscriptaddress,preprint,showpacs,showkeys,amsmath,amssymb]{revtex4}

\usepackage{graphicx}
\usepackage{dcolumn}
\usepackage{bm}

\begin{document}

\title{Coherent States and a Path Integral\\for the Relativistic Linear Singular Oscillator}

\author{S.M. Nagiyev}
\affiliation{%
Institute of Physics, Azerbaijan National Academy of Sciences\\
Javid av. 33, AZ1143, Baku, Azerbaijan
}%
\author{E.I. Jafarov}%
 \altaffiliation{Corresponding Author}
 \email{azhep@physics.ab.az}
\affiliation{%
Institute of Physics, Azerbaijan National Academy of Sciences\\
Javid av. 33, AZ1143, Baku, Azerbaijan
}%
\affiliation{Department of Applied Mathematics and Computer Science, Ghent University\\Krijgslaan 281-S9, B-9000 Gent, Belgium}

\author{M.Y. Efendiyev}
\affiliation{
Azerbaijan Cooperation University\\
Narimanov av. 86, AZ1106, Baku, Azerbaijan
}%

\date{\today}

\begin{abstract}
The $SU(1,1)$ coherent states for a relativistic model of the linear singular oscillator are considered. The corresponding partition function is evaluated. The path integral for the transition amplitude between $SU(1,1)$ coherent states is given. Classical equations of the motion in the generalized curved phase space are obtained. It is shown that the use of quasiclassical Bohr-Sommerfeld quantization rule yields the exact expression for the energy spectrum.
\end{abstract}


\maketitle

\section{Introduction}

Coherent States (CS) are a useful tool for studying quantum systems \cite{hassouni,shreecharan,wu}. The use of the CS makes it possible to apply more transparent classical language to describe the quantum phenomena \cite{fox,su}.

The concept of CS was first introduced for the boson oscillator \cite{glauber,klauder}. In this case they are closely related with the unitary representations of the Heisenberg-Weyl group. Later on, the generalized CS, associated with the unitary representations of an arbitrary Lie group, have been defined \cite{perelomov}. The notion of generalized CS arises, when we attempt to construct quasi-classical states for dynamical systems other than the harmonic oscillator \cite{chernyak,xu}.

In the present work the technique of constructing a path integral representation for the transition amplitude (propagator) between $SU(1,1)$ coherent states, developed in \cite{perelomov,kuratsuji,gerry1,gerry2,gerry3} is applied to the relativistic model of the linear singular oscillator \cite{nagiyev1}. The same problem for the relativistic model of the harmonic oscillator was considered in \cite{atakishiyev1}.

This paper has following structure: Section 2 presents a brief description of the relativistic linear singular oscillator and its $SU(1,1)$ dynamical symmetry group. The explicit form of $SU(1,1)$ CS for this problem is given and the corresponding partition function is evaluated in Section 3. In Section 4 we consider a path integral expression of the propagator in $SU(1,1)$ CS and examine the corresponding classical limit. It is shown that the use of the quasiclassical Bohr-Summerfield quantization rule yields the exact expression for the energy spectrum of the considered relativistic linear singular oscillator.

\section{The relativistic linear singular oscillator and $SU(1,1)$ dynamical symmetry group}

Recently, we constructed a relativistic model of the quantum linear singular oscillator \cite{nagiyev1}, which can be applied for studying relativistic physical systems as well as systems on a lattice. This model is formulated in the framework of the finite-difference relativistic quantum mechanics, which was developed in several papers and applied to the solution of a lot of problems in particle physics \cite{kadyshevsky1,kadyshevsky2,freeman,klein,atakishiyev2,atakishiyev3,atakishiyev4}.

The Hamiltonian of the relativistic model of the linear singular oscillator under consideration is a finite-difference operator \cite{nagiyev1}

\begin{equation}
\label{1}
H = mc^2 \left[ {\cosh i\partial _\rho   + \frac{1}{2}\omega _0 ^2 \rho ^{(2)} e^{i\partial _\rho  }  + \frac{{g_0 }}{{\rho ^{(2)} }}e^{i\partial _\rho  } } \right],
\end{equation}
where $\rho  = x/\mathchar'26\mkern-10mu\lambda$ is a dimensionless variable, $\mathchar'26\mkern-10mu\lambda  = \frac{\hbar }{{mc}}$ is the Compton wavelength of the particle, $\omega _0  = \frac{{\hbar \omega }}{{mc^2 }}$, $g_0  = \frac{{mg}}{\hbar }$, and $\rho ^{(2)}$ is the generalized degree \cite{nagiyev2}

$$
\rho ^{(\delta )}  = i^\delta  \frac{{\Gamma \left( {\delta  - i\rho } \right)}}{{\Gamma \left( { - i\rho } \right)}}.
$$

The eigenfunctions of the Hamiltonian (\ref{1}) in the interval $0 < \rho  < \infty $ are expressed in terms of the continuous dual Hahn polynomials $S_n \left( {x^2 ;a,b,c} \right)$, i.e. 

\begin{eqnarray}
\label{2}
 \psi _n \left( \rho  \right) = c_n \omega _0 ^{i\rho } \left( { - \rho } \right)^{\left( \alpha  \right)} \Gamma \left( {\nu  + i\rho } \right)S_n \left( {\rho ^2 ;\alpha ,\nu ,\frac{1}{2}} \right), \nonumber \\ \\
 c_n  = \sqrt {\frac{2}{{n!\Gamma \left( {n + \alpha  + \nu } \right)\Gamma \left( {n + \alpha  + 1/2} \right)\Gamma \left( {n + \nu  + 1/2} \right)}}} . \nonumber 
\end{eqnarray}

Here we have introduced the notations

\begin{eqnarray}
\label{3}
 \alpha  = \frac{1}{2} + \frac{1}{2}\sqrt {1 + \frac{2}{{\omega _0 ^2 }}\left( {1 - \sqrt {1 - 8g_0 \omega _0 ^2 } } \right)} , \nonumber \\ \\
 \nu  = \frac{1}{2} + \frac{1}{2}\sqrt {1 + \frac{2}{{\omega _0 ^2 }}\left( {1 + \sqrt {1 - 8g_0 \omega _0 ^2 } } \right)} . \nonumber 
\end{eqnarray}

The functions (\ref{2}) are orthonormal

\begin{equation}
\label{4}
\int\limits_0^\infty  {\psi _n ^* \left( \rho  \right)\psi _m \left( \rho  \right)d\rho }  = \delta _{nm} .
\end{equation}

A dynamical symmetry group for the model of the relativistic linear singular oscillator under consideration is the $SU(1,1)$ group. The corresponding Lie algebra is formed by the generators

\begin{equation}
\label{5}
K_0  = \frac{1}{{2\hbar \omega }}H,\quad K^ -   = A^ -  f^{ - 1} \left( H \right),\quad K^ +   = f^{ - 1} \left( H \right)A^ +  ,
\end{equation}
where

$$
f\left( H \right) = \left\{ {\left[ {\frac{H}{{mc^2 }} + \omega _0 \left( {\alpha  - \nu  - 1} \right)} \right]\left[ {\frac{H}{{mc^2 }} + \omega _0 \left( {\nu  - \alpha  - 1} \right)} \right]} \right\}^{1/2}. 
$$

Having defined a generalized momentum operator 

$$
P =  - mc\left[ {\sinh \left( {i\partial _\rho  } \right) + \frac{1}{2}\omega _0^2 \rho ^{(2)} e^{i\partial _\rho  }  + \frac{{g_0 }}{{\rho ^{(2)} }}e^{i\partial _\rho  } } \right]
$$
by means of the commutator

$$
\left[ {\rho ,H} \right] = icP,
$$
the operators $A^ \pm$ may be written as

\begin{equation}
\label{6}
A^ \pm   = \frac{1}{{2\omega _0 }}\left[ {\left( {\omega _0 \rho  \mp \frac{i}{{mc}}P} \right)^2  - \frac{{2g_0 }}{{\rho ^2  + 1}}} \right].
\end{equation}

The generators (\ref{5}) satisfy the commutation relations 

\begin{equation}
\label{7}
\left[ {K_0 ,K^ \pm  } \right] =  \pm K^ \pm  ,\quad \left[ {K^ -  ,K^ +  } \right] = 2K_0 .
\end{equation}

The operator $K_0 $ has a discrete spectrum in a infinite-dimensional unitary irreducible representation $D^ +  (k)$ such that 

\begin{equation}
\label{8}
K_0 \psi _n  = \left( {n + k} \right)\psi _n ,
\end{equation}
where $n = 0,1,2, \ldots $, and $k>0$. The Casimir invariant is 

$$
K^2  = K_0 ^2  - \frac{1}{2}\left( {K^ +  K^ -   + K^ -  K^ +  } \right) = k\left( {k - 1} \right)\hat I.
$$

For the operators (\ref{5}) one has $K^2  = \frac{{\alpha  + \nu }}{2}\left( {\frac{{\alpha  + \nu }}{2} - 1} \right)$, so that $k = \left( {\alpha  + \nu } \right)/2$. Thus from (\ref{5}) and (\ref{8}) we determine the energy levels as

\begin{equation}
\label{9}
E_n  = 2\hbar \omega \left( {n + k} \right) = \hbar \omega \left( {2n + \alpha  + \nu } \right).
\end{equation}

Let us emphasize that due to the commutation relations (\ref{7}) the action of the generators $K^ \pm  $ on the wavefunctions $\psi _n $ is given by

\begin{eqnarray}
\label{10}
 K^ -  \psi _n  = k_n \psi _{n - 1} ,\quad K^ +  \psi _n  = k_{n + 1} \psi _{n + 1} , \nonumber \\ \\
 k_n  = \sqrt {n(n + 2k - 1)}  = \sqrt {n(n + \alpha  + \nu  - 1)} . \nonumber
\end{eqnarray}

From (\ref{10}) follows that

\begin{eqnarray}
\label{11}
 \psi _n  = N_n \left( {K^ +  } \right)^n \psi _0 , \nonumber \\ 
 N_n ^{ - 1}  = k_1 k_2  \cdots k_n  = \sqrt {n!\left( {\alpha  + \nu } \right)_n } , \\ 
 \left( a \right)_n  = \Gamma \left( {n + a} \right)/\Gamma (a). \nonumber 
\end{eqnarray}

In the non-relativistic limit, when $c \to \infty $ the wave-functions $\psi _n \left( \rho  \right)$ coincide with the wavefunctions of the non-relativistic linear singular oscillator. In this limit we also have

\begin{eqnarray}
\label{12}
 \mathop {\lim }\limits_{c \to \infty } \left( {H - mc^2 } \right) = H_{nonrel}  = \hbar \omega \left( { - \frac{1}{2}\partial _\xi  ^2  + \frac{1}{2}\xi ^2  + \frac{{g_0 }}{{\xi ^2 }}} \right), \nonumber \\ 
 \mathop {\lim }\limits_{c \to \infty } \left( {E_n  - mc^2 } \right) = E_n ^{nonrel}  = \hbar \omega \left( {2n + d + 1} \right), \nonumber \\ 
 \mathop {\lim }\limits_{c \to \infty } \frac{1}{2}A^ -   = K^ -  _{nonrel}  = \frac{1}{2}\left( {\left( {a^ -  } \right)^2  - \frac{{g_0 }}{{\xi ^2 }}} \right), \nonumber \\ 
 \mathop {\lim }\limits_{c \to \infty } \frac{1}{2}A^ +   = K^ +  _{nonrel}  = \frac{1}{2}\left( {\left( {a^ +  } \right)^2  - \frac{{g_0 }}{{\xi ^2 }}} \right), \\ 
 \mathop {\lim }\limits_{c \to \infty } \Pi  =  - i\sqrt {m\hbar \omega } \partial _\xi   =  - i\hbar \partial _x  = p_x , \nonumber \\ 
 \mathop {\lim }\limits_{c \to \infty } \alpha  = d + 1/2, \nonumber \\ 
 \mathop {\lim }\limits_{c \to \infty } \left( {\nu  - \mu } \right) = 1/2, \nonumber 
\end{eqnarray}
where $d = \frac{1}{2}\sqrt {1 + 8g_0 } $, $\xi  = \sqrt {\frac{{m\omega }}{\hbar }} x$ and

$$
a^ +   = \frac{1}{{\sqrt 2 }}\left( {\xi  - \partial _\xi  } \right),\quad a^ +   = \frac{1}{{\sqrt 2 }}\left( {\xi  - \partial _\xi  } \right)
$$
are the usual creation and annihilation operators.

\section{$SU(1,1)$ Coherent States}

$SU(1,1)$ CS $\left| {\zeta ,k} \right\rangle $ are defined by acting with the displacement operator $D\left( \beta  \right) = \exp \left( {\beta K^ +   - \beta ^* K^ -  } \right)$ on the ground state wavefunctions $\psi _0 \left( \rho  \right)$, i.e.
\begin{equation}
\label{13}
\left| {\zeta ,k} \right\rangle  = D\left( \beta  \right)\psi _0 \left( \rho  \right) = \left( {1 - \left| \zeta  \right|^2 } \right)^k e^{\zeta K^ +  } \psi _0 \left( \rho  \right),
\end{equation}
where $\beta  =  - \frac{\tau }{2}e^{ - i\varphi } $ and $\zeta  =  - \tanh \frac{\tau }{2}e^{ - i\varphi } $, $\tau$ and $\varphi $ are group parameters. From (\ref{10}) and (\ref{13}) it follows that the decomposition of $\left| {\zeta ,k} \right\rangle $ over the wavefunctions   $\psi _n \left( \rho  \right)$ (\ref{2}) has the form

\begin{equation}
\label{14}
\left| {\zeta ,k} \right\rangle  = \left( {1 - \left| \zeta  \right|^2 } \right)^k \sum\limits_{n = 0}^\infty  {\sqrt {\frac{{\left( {2k} \right)_n }}{{n!}}} \zeta ^n \psi _n \left( \rho  \right)} .
\end{equation}

Using (\ref{2}) one can rewrite (\ref{14}) as follows

\begin{eqnarray}
 \left| {\zeta ,k} \right\rangle  = \left( {1 - \left| \zeta  \right|^2 } \right)^k \sqrt {\frac{2}{{\Gamma \left( {\alpha  + \nu } \right)\Gamma \left( {\alpha  + 1/2} \right)\Gamma \left( {\nu  + 1/2} \right)}}} \omega _0 ^{i\rho } \left( { - \rho } \right)^{\left( \alpha  \right)} \Gamma \left( {\nu  + i\rho } \right) \times  \nonumber \\ \label{14a} \\ 
 \sum\limits_{n = 0}^\infty  {\frac{{\zeta ^n }}{{n!}}\left[ {\left( {\alpha  + 1/2} \right)_n \left( {\nu  + 1/2} \right)_n } \right]^{ - 1/2} S_n \left( {\rho ^2 ;\alpha ,\nu ,\frac{1}{2}} \right)} . \nonumber
\end{eqnarray}

One can look for the explicit expression of CS (\ref{14a}) taking into account Hermiticity conditions of the Hamiltonian. Hermiticity condition of the Hamiltonian imposes a restriction on the values of the quantity $g_0$. Therefore, eigenvalues (\ref{9}) are real only in case when $\alpha$ and $\nu$ are real or complex-conjugate. We will calculate series (\ref{14a}) for the case, when $\alpha$ and $\nu$ are equal or complex-conjugate, which imposes the condition $g_0  \ge \frac{1}{{8\omega _0 ^2 }}$. The behavior of $\alpha$ and $\nu$ are presented in Figs. 1 and 2.

\begin{figure*}
\includegraphics{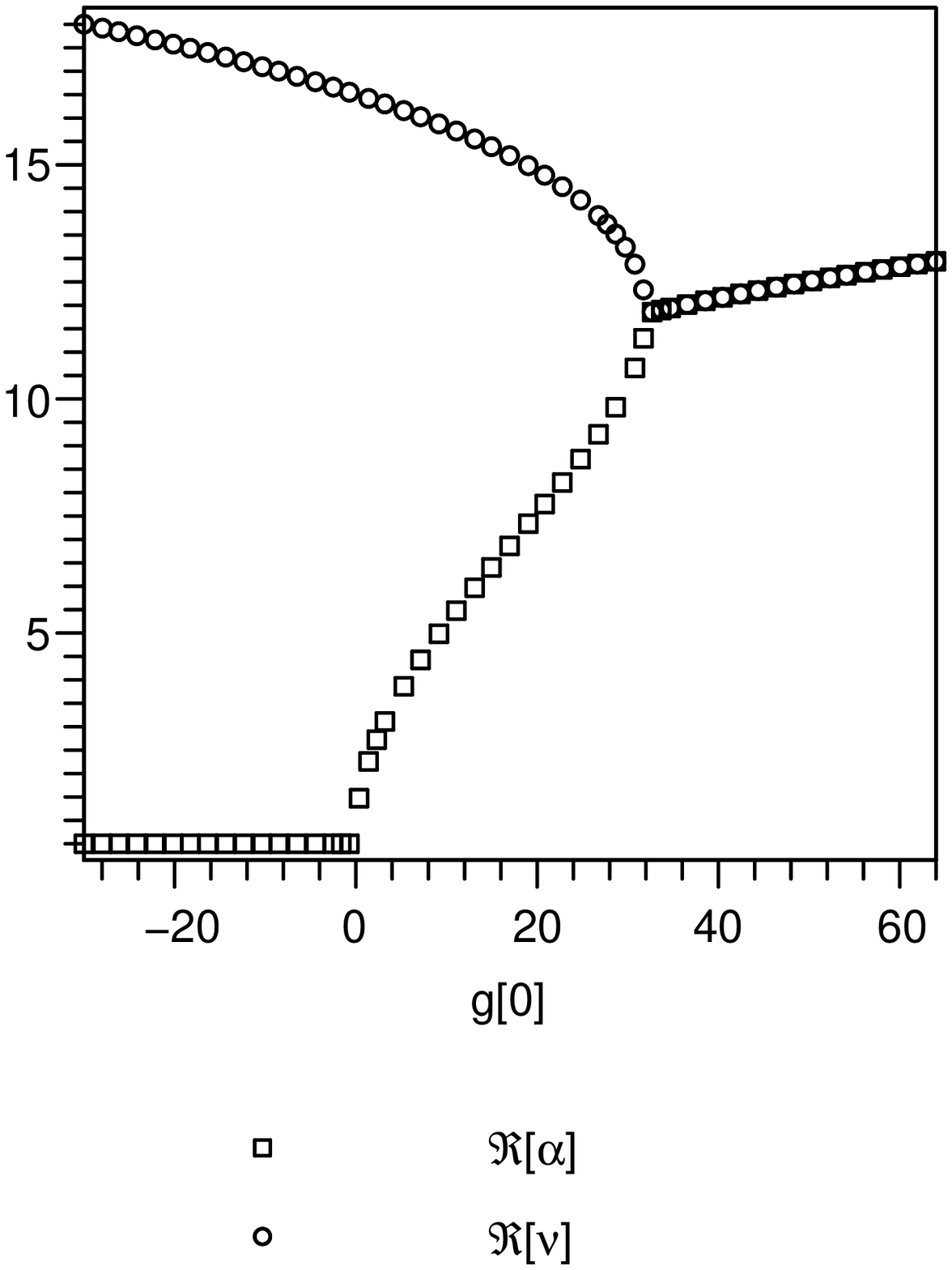}
\caption{The behavior of $\alpha$ and $\nu$: real parts}
\end{figure*}

\begin{figure*}
\includegraphics{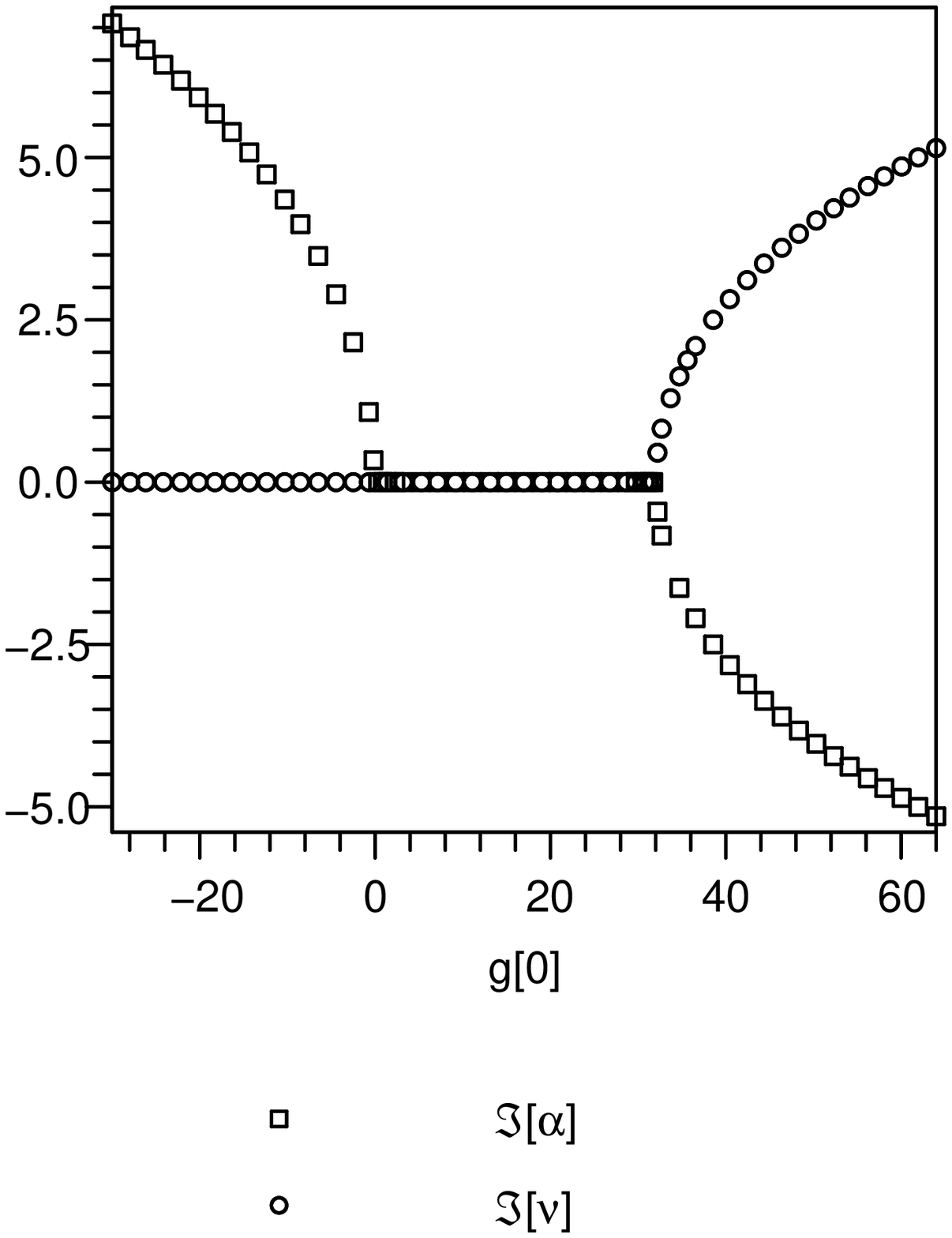}
\caption{The behavior of $\alpha$ and $\nu$: imaginary parts}
\end{figure*}

Mentioned above condition allows us to rewrite (\ref{14a}) as

\begin{eqnarray}
 \left| {\zeta ,k} \right\rangle  = \left( {1 - \left| \zeta  \right|^2 } \right)^k \sqrt {\frac{2}{{\Gamma \left( {\alpha  + \nu } \right)\Gamma \left( {\alpha  + 1/2} \right)\Gamma \left( {\nu  + 1/2} \right)}}} \omega _0 ^{i\rho } \left( { - \rho } \right)^{\left( \alpha  \right)} \Gamma \left( {\nu  + i\rho } \right) \times  \nonumber \\ \label{14b} \\ 
 \sum\limits_{n = 0}^\infty  {\frac{{\zeta ^n }}{{n!\left( {\left| \alpha  \right| + 1/2} \right)_n }}S_n \left( {\rho ^2 ;\left| \alpha  \right|,\left| \alpha  \right|,\frac{1}{2}} \right)} . \nonumber
\end{eqnarray}

By the use of the following generation function for the continuous dual Hahn polynomials \cite{koekoek}
$$
\sum\limits_{n = 0}^\infty  {\frac{{S_n \left( {x^2 ;a,b,c} \right)}}{{(a + c)_n n!}}t^n }  = (1 - t)^{ - b + ix} {}_2F_1 \left( {\left. {\begin{array}{*{20}c}
   {a + ix,c + ix}  \\
   {a + c}  \\
\end{array}} \right|t} \right)
$$
one can simplify (\ref{14b}) as

\begin{eqnarray}
 \left| {\zeta ,k} \right\rangle  = \left( {1 - \left| \zeta  \right|^2 } \right)^k \sqrt {\frac{2}{{\Gamma \left( {\alpha  + \nu } \right)\Gamma \left( {\alpha  + 1/2} \right)\Gamma \left( {\nu  + 1/2} \right)}}} \omega _0 ^{i\rho } \left( { - \rho } \right)^{\left( \alpha  \right)} \Gamma \left( {\nu  + i\rho } \right) \times  \nonumber \\ \label{14c} \\ 
 (1 - \zeta )^{ - \left| \alpha  \right| + i\rho } {}_2F_1 \left( {\left. {\begin{array}{*{20}c}
   {\left| \alpha  \right| + i\rho ,\frac{1}{2} + i\rho }  \\
   {\left| \alpha  \right| + \frac{1}{2}}  \\
\end{array}} \right|\zeta } \right). \nonumber 
\end{eqnarray}

The $SU(1,1)$ CS (\ref{13}) are orthogonal and the overlap of two states $\left| {\zeta ,k} \right\rangle $ and $\left| {\zeta' ,k} \right\rangle $ is given as

\begin{equation}
\label{15}
\left\langle {{\zeta ',k}}
 \mathrel{\left | {\vphantom {{\zeta ',k} {\zeta ,k}}}
 \right. \kern-\nulldelimiterspace}
 {{\zeta ,k}} \right\rangle  = \left( {1 - \left| {\zeta '} \right|^2 } \right)^k \left( {1 - \left| \zeta  \right|^2 } \right)^k \left( {1 - \zeta '^* \zeta } \right)^{ - 2k} .
\end{equation}

The important property of these states is the completeness relation

\begin{equation}
\label{16}
\int {d\mu _k (\zeta )\left| {\zeta ,k} \right\rangle \left\langle {\zeta ',k} \right|}  = 1,
\end{equation}
where
\begin{equation}
\label{17}
d\mu _k \left( \zeta  \right) = \frac{{2k - 1}}{\pi }\frac{{d^2 \zeta }}{{\left( {1 - \left| \zeta  \right|^2 } \right)^2 }}.
\end{equation}

The matrix elements of the generators $K^-$, $K^+$, $K_0$ in the $SU(1,1)$ CS have the form

\begin{eqnarray}
\label{18}
 \left\langle {\zeta ',k} \right|K^ -  \left| {\zeta ,k} \right\rangle  = \frac{{2k\zeta }}{{1 - \zeta '^* \zeta }}\left\langle {{\zeta ',k}}
 \mathrel{\left | {\vphantom {{\zeta ',k} {\zeta ,k}}}
 \right. \kern-\nulldelimiterspace}
 {{\zeta ,k}} \right\rangle , \nonumber \\ 
 \left\langle {\zeta ',k} \right|K^ +  \left| {\zeta ,k} \right\rangle  = \frac{{2k\zeta '^* }}{{1 - \zeta '^* \zeta }}\left\langle {{\zeta ',k}}
 \mathrel{\left | {\vphantom {{\zeta ',k} {\zeta ,k}}}
 \right. \kern-\nulldelimiterspace}
 {{\zeta ,k}} \right\rangle , \\ 
 \left\langle {\zeta ',k} \right|K_0 \left| {\zeta ,k} \right\rangle  = \frac{{k\left( {1 + \zeta '^* \zeta } \right)}}{{1 - \zeta '^* \zeta }}\left\langle {{\zeta ',k}}
 \mathrel{\left | {\vphantom {{\zeta ',k} {\zeta ,k}}}
 \right. \kern-\nulldelimiterspace}
 {{\zeta ,k}} \right\rangle . \nonumber 
 \end{eqnarray}

The transition amplitude (propagator) between $SU(1,1)$ CS is defined as

\begin{eqnarray*}
K\left( {\zeta ';\zeta ;T} \right) &=& \left\langle {\zeta ';k} \right|\exp \left( { - \frac{i}{\hbar }TH} \right)\left| {\zeta ;k} \right\rangle , \\
 &=& \left\langle {\zeta ';k} \right|\exp \left[ { - 2i\omega _0 TK_0 } \right]\left| {\zeta ;k} \right\rangle .
\end{eqnarray*}

Using (\ref{14}) and (\ref{15}) it is easy to show that

\begin{equation}
\label{19}
K\left( {\zeta ',\zeta ,T} \right) = e^{ - 2i\omega kT} \frac{{\left( {1 - \left| \zeta  \right|^2 } \right)^k \left( {1 - \left| {\zeta '} \right|^2 } \right)^k }}{{\left( {1 - \zeta '^* \zeta e^{ - 2i\omega kT} } \right)^{2k} }}.
\end{equation}

The partition function for the relativistic model of the linear singular oscillator under consideration is given as

\begin{eqnarray*}
Z_{rel} = {\mathop{\rm Tr}\nolimits}\;K\left( {\zeta ,\zeta '; - i\hbar \beta } \right) &=& \frac{{e^{ - 2k\hbar \omega \beta } }}{{1 - e^{ - 2k\hbar \omega \beta } }} \\
 &=& e^{ - 2\hbar \omega \left( {\alpha  + \nu  - d - 1} \right)} Z_{nonrel} ,
\end{eqnarray*}
where $Z_{nonrel}$ is the partition function for the nonrelativistic linear singular oscillator.

\section{Path integral and classical equations of motion in the generalized path space}

Following the paper \cite{kuratsuji,gerry1} we now derive the path integral expression for the amplitude (\ref{19}). Defining $\varepsilon  = T/N$ and using the completeness relation (\ref{16}) it is possible to represent (\ref{19}) as

\begin{equation}
\label{20}
K\left( {\zeta ',\zeta ;T} \right) = \int  \cdots  \int {\prod\limits_{j = 1}^{N - 1} {d\mu _k \left( {\zeta _j } \right)\left\langle {\zeta ',k} \right|e^{ - \frac{i}{\hbar }\varepsilon H} \left| {\zeta _{N - 1} ,k} \right\rangle \left\langle {\zeta _{N - 1} ,k} \right|e^{ - \frac{i}{\hbar }\varepsilon H} \left| {\zeta _{N - 2} ,k} \right\rangle  \cdots } } \left\langle {\zeta _1 ,k} \right|e^{ - \frac{i}{\hbar }\varepsilon H} \left| {\zeta ,k} \right\rangle .
\end{equation}

With the help of (\ref{18}) it is easy to show that for small $\varepsilon $ each factor in the integrand (\ref{20}) can be written as

$$
\left\langle {\zeta _j ,k} \right|e^{ - \frac{i}{\hbar }\varepsilon H} \left| {\zeta _{j - 1} ,k} \right\rangle  \cong \exp \left[ {\ln \left\langle {{\zeta _j ,k}}
 \mathrel{\left | {\vphantom {{\zeta _j ,k} {\zeta _{j - 1} ,k}}}
 \right. \kern-\nulldelimiterspace}
 {{\zeta _{j - 1} ,k}} \right\rangle } \right] - \frac{{i\varepsilon }}{\hbar }H_k \left( {\zeta _j ^* ,\zeta _{j - 1} } \right),
$$
where

\begin{equation}
\label{21}
H_k \left( {\zeta _j ^* ,\zeta _{j - 1} } \right) = \frac{{\left\langle {\zeta _j ,k} \right|H\left| {\zeta _{j - 1} ,k} \right\rangle }}{{\left\langle {{\zeta _j ,k}}
 \mathrel{\left | {\vphantom {{\zeta _j ,k} {\zeta _{j - 1} ,k}}}
 \right. \kern-\nulldelimiterspace}
 {{\zeta _{j - 1} ,k}} \right\rangle }} = 2k\hbar \omega \frac{{1 + \zeta _j ^* \zeta _{j - 1} }}{{1 - \zeta _j ^* \zeta _{j - 1} }}.
\end{equation}

If we take into account (\ref{15}) and fact that $\zeta _{j - 1}  = \zeta _j  - \Delta \zeta _j $, we can write

$$
\ln \left\langle {{\zeta _j ,k}}
 \mathrel{\left | {\vphantom {{\zeta _j ,k} {\zeta _{j - 1} ,k}}}
 \right. \kern-\nulldelimiterspace}
 {{\zeta _{j - 1} ,k}} \right\rangle  \cong \frac{k}{{1 - \left| \zeta  \right|^2 }}\left( {\zeta _j \Delta \zeta _j ^*  - \zeta _j ^* \Delta \zeta _j } \right).
$$

Thus, when $N \to \infty $ (or $\varepsilon  \to 0$) we arrive at the following path integral for the amplitude (\ref{19})

\begin{equation}
\label{22}
K\left( {\zeta ',\zeta ;T} \right) = \int {D\mu _k \left( \zeta  \right)\exp \left\{ {\frac{i}{\hbar }\int\limits_0^T {L_k \left( {\zeta ,\dot \zeta ,\zeta ^* ,\dot \zeta ^* } \right)dt} } \right\}} ,
\end{equation}
with the classical Lagrangian

\begin{equation}
\label{23}
L_k \left( {\zeta ,\dot \zeta ,\zeta ^* ,\dot \zeta ^* } \right) = \frac{{i\hbar k}}{{1 - \left| \zeta  \right|^2 }}\left( {\zeta ^* \dot \zeta  - \zeta \dot \zeta ^* } \right) - H_k \left( {\zeta ^* ,\zeta } \right)
\end{equation}
in a generalized curved phase space in the form of a Lobachevsky plane.

The corresponding classical Euler-Lagrange equations have the form

\begin{equation}
\label{24}
\frac{d}{{dt}}\left( {\frac{{\partial L_k }}{{\partial \dot \zeta }}} \right) = \frac{{\partial L_k }}{{\partial \zeta }},\quad \frac{d}{{dt}}\left( {\frac{{\partial L_k }}{{\partial \dot \zeta ^* }}} \right) = \frac{{\partial L_k }}{{\partial \zeta ^* }}.
\end{equation}

Using (\ref{23}) we can represent (\ref{24}) in the form of Hamiltonian's equations:

\begin{equation}
\label{25}
\dot \zeta  = \frac{{\left( {1 - \left| \zeta  \right|^2 } \right)^2 }}{{2i\hbar k}}\frac{{\partial H_k }}{{\partial \zeta ^* }},\quad \dot \zeta ^*  = \frac{{\left( {1 - \left| \zeta  \right|^2 } \right)^2 }}{{2i\hbar k}}\frac{{\partial H_k }}{{\partial \zeta }}.
\end{equation}

If we define a Poisson bracket by

\begin{equation}
\label{26}
\left\{ {A,B} \right\} = \frac{{\left( {1 - \left| \zeta  \right|^2 } \right)^2 }}{{2i\hbar k}}\left( {\frac{{\partial A}}{{\partial \zeta }}\frac{{\partial B}}{{\partial \zeta ^* }} - \frac{{\partial A}}{{\partial \zeta ^* }}\frac{{\partial B}}{{\partial \zeta }}} \right),
\end{equation}
then we can write the equations (\ref{25}) in a more compact form as

\begin{equation}
\label{27}
\dot \zeta  = \left\{ {\zeta ,H_k } \right\},\quad \dot \zeta ^*  = \left\{ {\zeta ^* ,H_k } \right\}.
\end{equation}

Since in our case $H_k \left( {\zeta ^* ,\zeta } \right) \equiv H_k \left( \tau  \right) = 2\hbar \omega k\cosh \left( \tau  \right)$, the equations (\ref{27}) written in terms of the group parameters $\tau$ and $\varphi $ will be reduced to

\begin{equation}
\label{28}
\dot \tau  = \left\{ {\tau ,H_k \left( \tau  \right)} \right\} = 0,\quad \dot \varphi  = \left\{ {\varphi ,H_k \left( \tau  \right)} \right\} = 2\omega .
\end{equation}

The solutions of (\ref{28}) are $\tau  = const$ and $\varphi  = 2\omega t\varphi _0 $. Therefore, the classical motion in the curved phase space is oscillator like.
 
In terms of $\tau$ and $\varphi $ the Lagrangian (\ref{23}) becomes

\begin{equation}
\label{29}
L_k  = \hbar k\left[ {\left( {\cosh \left( \tau  \right) - 1} \right)\dot \varphi  - 2\omega \cosh \left( \tau  \right)} \right] \equiv \hbar k\tilde L\left( {\tau ,\varphi } \right).
\end{equation}

Using the momentum $p = \partial \tilde L/\partial \dot \varphi  = \cosh \left( \tau  \right) - 1$, canonically conjugate to the coordinate $\varphi $ we may write

\begin{equation}
\label{30}
\tilde L\left( {p,\varphi } \right) \equiv \tilde L\left( {\tau ,\varphi } \right) = p\dot \varphi  - 2\omega \left( {p + 1} \right).
\end{equation}

Substituting (\ref{30}) into (\ref{22}) we arrive at the path-integral expression

\begin{equation}
\label{31}
K\left( {\zeta ',\zeta ;T} \right) = \int {D\mu _k \left( {p,\varphi } \right)\exp \left\{ {ik\int\limits_0^T {\tilde L\left( {p,\varphi } \right)dt} } \right\}} .
\end{equation}

Since in the $\hbar  \to 0$ limit the parameter $k = \left( {\alpha  + \nu } \right)/2$ characterizing the irreducible representation $D^ +  \left( k \right)$ of the dynamical symmetry group $SU(1,1)$ behaves as $k \cong \frac{{mc^2 }}{{\hbar \omega }}$, from (\ref{31}) it follows that for $k$ sufficiently large, the motion of the relativistic linear singular oscillator in the curved phase space becomes quasiclassical.

Thus, when $k \to \infty $ we can use Bohr-Sommerfeld quantization rule to find the energy spectrum $E_k  = H_k \left( \tau  \right)$, i.e.

\begin{equation}
\label{32}
\oint {pd\varphi  = \frac{{2\pi }}{k}n} ,\;n = 0,1,2, \ldots \quad .
\end{equation}

From (\ref{32}) follows that $p = n/k$ and therefore

\begin{eqnarray}
\label{33}
E_k  = H_k \left( \tau  \right) &=& 2\hbar \omega k\cosh (\tau ) = 2\hbar \omega k(p + 1) \nonumber \\
 &=& 2\hbar \omega (n + k) = \hbar \omega (2n + \alpha  + \nu ).
\end{eqnarray}

Therefore, as in the non-relativistic case, the Bohr-Sommerfeld quantization rule yields for the energy spectrum of the relativistic linear singular oscillator the exact expression (\ref{33}).

\section{Conclusion}

In spite of many attractive papers devoted to construction of CS for non-relativistic quantum systems, the number of works studying relativistic approaches to CS and path integral formulation of the quantum systems is still rather few \cite{atakishiyev1,bagrov,lev,haghlighat}.

In this paper we have considered the   CS for a relativistic model of the linear singular oscillator and obtained their explicit form. Thereafter, a path integral expression of the transition amplitude between   CS has been studied and corresponding classical limits are shown. By the use of path integral approach the classical equations of the motion in the generalized curved phase space are obtained. It was shown that the use of quasiclassical Bohr-Sommerfeld quantization rule yields the exact expression for the energy spectrum.

\section*{Acknowledgement}

One of the authors (E.I.J.) would like to acknowledge that this work is performed in the framework of the Fellowship 05-113-5406 under the INTAS-Azerbaijan YS Collaborative Call 2005.


\end{document}